\documentclass[conference]{IEEEtran}
\usepackage{amsmath, graphicx}
\usepackage{subcaption, siunitx, booktabs}
\usepackage{bm, amsfonts}
\usepackage{cite}
\usepackage{soul}

\def\x{{\bm x}}
\def\y{{\bm y}}
\def\w{{\bm w}}

\def\T{{\mathsf T}}

\title{Anomalous Sound Detection Based on Machine Activity Detection}
\author{
    \IEEEauthorblockN{
        Tomoya Nishida, Kota Dohi, Takashi Endo, Masaaki Yamamoto, Yohei Kawaguchi
        }
    \IEEEauthorblockA{
        Research and Development Group, Hitachi, Ltd.,\\
        1-280, Higashi-koigakubo, Kokubunji-shi, Tokyo 185-8601, Japan
        }
    }
\begin{document}
%\ninept
%
\maketitle
\begin{abstract}
We have developed an unsupervised anomalous sound detection method for machine condition monitoring that utilizes an auxiliary task --- detecting when the target machine is active. First, we train a model that detects machine activity by using normal data with machine activity labels and then use the activity-detection error as the anomaly score for a given sound clip if we have access to the ground-truth activity labels in the inference phase. If these labels are not available, the anomaly score is calculated through outlier detection on the embedding vectors obtained by the activity-detection model. Solving this auxiliary task enables the model to learn the difference between the target machine sounds and similar background noise, which makes it possible to identify small deviations in the target sounds. Experimental results showed that the proposed method improves the anomaly-detection performance of the conventional method complementarily by means of an ensemble.

\end{abstract}
\begin{IEEEkeywords}
Machine health monitoring, anomalous sound detection, self-supervised learning, machine activity detection
\end{IEEEkeywords}
\section{INTRODUCTION}
\label{sec:intro}
\vspace{-5pt}
Anomalous sound detection (ASD) is a task to identify whether a given sound is normal or anomalous. 
Since mechanical failure often causes machines to emit anomalous sounds, ASD has attracted attention for its application to machine condition monitoring~\cite{koizumi2019unsupervised}, an essential technology for artificial intelligence-based factory automation.
ASD is typically conducted in an unsupervised manner, meaning that only normal sounds are used for training. 
This is because anomalous sounds occur in rare situations and are highly diverse, making them almost impossible to collect. 
Most methods for unsupervised ASD~(UASD) first learn a model of the collected normal sounds ~\cite{koizumi2019batch, kawachi2018complementary, yamaguchi2019adaflow, marchi2015novel, hayashi2020conformer, harsh2020dagmm}. 
They then calculate the anomaly score of an observed sound on the basis of how well the sound fits the learned model. 
The sound is identified as anomalous if the score exceeds a preset threshold.

UASD for machine condition monitoring is often conducted in factories under noisy conditions, where the environmental noise tends to degrade the performance since the difference between normal and anomalous sounds is relatively small. This phenomenon is more pronounced when the environmental noise is similar to the target machine sounds.
Recent methods for solving the noise problem~\cite{giri2020unsupervised, primus2020reframing, inoue2020detection, lopez2021ensemble, morita2021anomalous, wilkinghoff2021subcluster}. 
have utilized models that classify the sounds of the target machine and those of other similar machines, in contrast to the conventional UASD methods, which enables them to distinguish minor deviations between normal and anomalous sounds. An ensemble of these methods with other UASD methods has exhibited a good performance. 
However, it is extremely labor-intensive to find other machines similar to the target machine and then to record those sounds as training data in practical situations.

In this paper, we propose a UASD method that does not require sounds of other machines similar to the target machine.
To solve the UASD task, the proposed method utilizes a model trained to solve an auxiliary task of detecting when the target machine is active.
First, we train an activity-detection model that estimates when the target machine is active. 
Then, in the inference phase, we calculate the anomaly score by using the activity-detection error of the activity-detection model.
Since the activity-detection model is trained to distinguish the sounds of the target machine from environmental noise, it can detect anomalous sounds especially when environmental noise is similar to the sounds of the target machine.
Moreover, to enable inference without ground-truth machine activity labels, we propose applying an outlier detection method to the embeddings extracted from the activity-detection model.

\section{RELATED WORK}
\label{sec:related}
\vspace{-5pt}

\subsection{UASD methods}
\label{ssec:unsupervised}
\vspace{-3pt}
Various UASD methods have targeted machine condition monitoring. Most of them learn a model of the normal sounds and then detect sounds that deviate from the learned model as anomalous. Several models have been used for learning the normal model, including autoencoders (AEs)~\cite{koizumi2019batch}, variational autoencoders~\cite{kawachi2018complementary}, long short-term memories~\cite{marchi2015novel}, transformers~\cite{hayashi2020conformer}, normalizing flows~\cite{yamaguchi2019adaflow}, and Gaussian mixture models~(GMMs)~\cite{harsh2020dagmm}. With these models, the anomaly score $\mathcal{A}(\x; \bm{\theta})$ for a given input $\x$ is calculated in the inference phase, where $\bm{\theta}$ is the parameter of the model. If $\mathcal{A}(\x; \bm{\theta})$ exceeds a predefined threshold, input $\x$ is detected as anomalous. Extensions to AEs and VAEs have also been proposed, such as changing the reconstruction task to an interpolation task~\cite{suefusa2020anomalous}, which improves the detection performance for non-stationary sounds, or cascading various types of dereverberation methods before the model~\cite{kawaguchi2019anomaly}. All of these methods except the final one~\cite{kawaguchi2019anomaly} learn the normal model without distinguishing the target machine sounds from environmental noise, and as a result, the existence of environmental noise degrades the ASD performance.

\subsection{Self-supervised classification-based ASD methods}
\label{ssec:self-supervised}
\vspace{-3pt}
In Task 2 of Challenges on Detection and Classification of Acoustic Scenes and Events (DCASE Challenge) 2020 and 2021~\cite{koizumi2020description, kawaguchi2021description}, many methods based on learning classification models that can be interpreted as a variation of outlier exposure~\cite{hendrycks2019deep} performed well. These methods train a model that identifies the machine ID for a given audio clip (sounds of several different individuals were provided for each machine type, with the individuals tagged as IDs), and the classification error is used as the anomaly score~\cite{giri2020unsupervised, primus2020reframing, inoue2020detection,  lopez2021ensemble, morita2021anomalous, wilkinghoff2021subcluster}. 
A normalizing-flow-based method using data from multiple machine IDs has also been proposed~\cite{dohi2021flow-based}.
These methods tend to achieve better performances than unsupervised methods thanks to learning a good decision boundary between normal and anomalous samples by using the sounds from other machine IDs as proxy outliers. However, they need the sound of each class of the classification task to be similar to achieve these results~\cite{primus2020anomalous, dohi2021flow-based}. While this is possible in competitions where sounds for multiple machine IDs are provided for each machine type, in practical use it is quite costly to find appropriate machines and record their sounds.

\section{PROBLEM STATEMENT}
\label{sec:prob}
\vspace{-5pt}

In this paper, we tackle the UASD task, i.e., anomalous sound detection under the condition that only normal sounds are available in the training phase. Unlike DCASE 2020 and 2021 Challenge Task 2, we consider a case where sounds for different individuals of the same machine type are unavailable.
We also make the following assumptions.
\begin{enumerate}
\item The target machine repeatedly starts and stops during sound recording. 
We call the time when the machine is running \textit{active} and the time when it is stopped \textit{inactive}. 
During the active time, both the target machine sound and environmental noise are recorded, while during the inactive time, only the noise is recorded.
\item The training data contains information about when the target machine started and stopped running (called \textit{activity labels}). 
If machine activity can be automatically recorded, the activity labels will be available in the training and inference phases. 
Even if they are not automatically recorded, the activity labels are available in the training phase because they can be annotated by hand.
\end{enumerate}

\section{PROPOSED METHOD}
\label{sec:prop}
\vspace{-5pt}
\subsection{Basic concept}
\vspace{-5pt}
The basic concept of the proposed UASD method is to use a model trained to solve an auxiliary task of detecting when the target machine is active.
We call this model the ``activity-detection model'' and train it by using normal sounds of the target machine with ground-truth activity labels. 
If the activity-detection model fails to detect the active time frames in the inference phase, we regard the sound clip as anomalous.
The activity-detection model is expected to learn a good decision boundary between the normal sounds of the target machine and other sounds, including anomalous sounds. The proposed UASD method based on activity detection works especially well when the environmental noise is similar to the target-machine sound. 
This case is likely to occur in factories because many machines are often similar to the target machine in operation.

\subsection{Case in which activity labels are available in inference}
%\vspace{-5pt}
\subsubsection{UASD based on supervised activity detection~(UASD-SAD)}
\label{ssec:activity detection}
\vspace{-5pt}
\begin{figure}[!t]
	\centering
	\includegraphics[width=7.0cm]{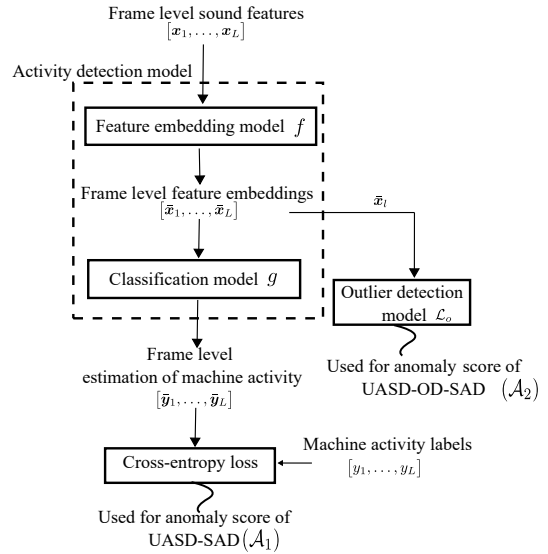}
	\caption{Overview of proposed UASD-SAD and UASD-OD-SAD.}
	\label{fig:overview}
	\vspace{-18pt}
\end{figure}

As shown in Fig.~\ref{fig:overview}, the activity-detection model consists of two components: a feature embedding model and a frame-wise classification model. First, $F$-dimensional frame-wise feature vectors (such as log-mel spectrograms) are extracted from sound clips and $L$ consecutive features $\begin{bmatrix}\x_1, \dots, \x_L \end{bmatrix},$ $\x_l \in \mathbb{R}^F~(l=1,\ldots, L)$ are taken as input. The task is to estimate the activity labels $[y_{1},\ldots,y_{L}]$, where $y_{l}=1$ if the target machine was active in the $l$-th time frame and $y_l=0$ if not. 
Note that only a few feature frames are input into this model, not the entire audio clip. 
The aim is to make activity detection difficult enough to be an auxiliary task for anomaly detection. 
If we input the whole audio clip, the activity will be detected precisely for both normal and anomalous sounds, and anomaly detection will not work.

The feature embedding model extracts $L$ embedding vectors $\begin{bmatrix}\bar{\x}_1,\cdots, \bar{\x}_L\end{bmatrix}$ from input as
\begin{align}
	\begin{bmatrix} \bar{\x}_{1}, \ldots, \bar{\x}_{L} \end{bmatrix}
		&= f
			\left(
				\x_{1},\ldots,\x_{L} ; \bm{\theta}_f
			\right),
\end{align}
where $\bm{\theta}_f$ denotes the parameters of the feature embedding model $f$. This model can be a convolutional neural network (CNN), a gated recurrent unit, or any other appropriate model.
The classification model $g$ then estimates the activity label of each frame from the feature embeddings, as
\begin{align}
	\bar{\y}_l = g(\bar{\x_l} ; \bm{\theta}_g),~l=1,\ldots,L,
\end{align}
where $\bar{\y}_l \in (0,1)^2 $ and $\bm{\theta}_g$ denotes the parameters of the classification model $g$. The first and second component of $\bar{\y}_l$ can be interpreted as the posterior probability of the $l$-th time frame being inactive and active, respectively. Typically, $g$ can be given as a combination of a linear transform and a softmax function as
\begin{align}
	g(\bar{\x}; \bm{\theta}_g) = 
		\begin{bmatrix}
			\frac{
				\exp\left({\w_1^\T \bar{\x}}\right)
				}
				{
				\exp\left({\w_1^\T \bar{\x}}\right) + \exp\left({\w_2^\T \bar{\x}}\right)
				} \\
			\frac{
				\exp\left({\w_2^\T \bar{\x}}\right)
				}
				{
				\exp\left({\w_1^\T \bar{\x}}\right) + \exp\left({\w_2^\T \bar{\x}}\right)
				}
		\end{bmatrix}, \label{eq:classifier}
\end{align}
where $\bm{\theta}_g = \{\w_1, \w_2\}$. Finally, the detection error is calculated as the cross entropy loss between $\begin{bmatrix}\bar{\y}_1, \dots, \bar{\y}_L \end{bmatrix}$ and the machine activity labels as
\begin{align}
	\mathcal{L}\left(
		\begin{bmatrix}\x_1, \dots, \x_L \end{bmatrix}; \bm{\theta}
	\right)
	 = -\sum_{l=1}^{L} 
		\log{
			\left(
				\left[  \bar{\y}_l \right]_{y_l + 1} 
			\right)
			},
\end{align}
where $\bm{\theta}=\{ \bm{\theta}_f, \bm{\theta}_g \}$ and $[\y]_i$ denotes the $i$-th component of $\y$. The detection error is used both as the cost function used for training the model and as the anomaly score for an obtained sound. Since this method utilizes a supervised learning task of activity detection, we refer to it as \textit{UASD based on supervised activity detection~(UASD-SAD)}.

\subsubsection{Overall cost function and anomaly score for a sound clip}
\label{ssec:}
We describe the overall cost function for a given training dataset and the anomaly score for a given sound clip. Both are calculated using a sliding window to extract $L$ consecutive frames of feature vectors, which are then used as the input of the activity-detection model.

Assume that the training data consists of $K$ sound clips of normal data $\mathcal{D}=\{ \bm{X}^{(k)} \}_{k=1}^{K}$, where each sound clips consists of $T_k(\geq L)$ frames of feature vectors: $\bm{X}^{(k)} = \begin{bmatrix} \x_1^{(k)},\ldots, \x_{T_k}^{(k)} \end{bmatrix}$. For the input of the activity-detection model, $L$ consecutive feature vectors starting from index $t$ are denoted as
\begin{align}
	\bm{X}^{(k)}_t = 
	\begin{bmatrix}
		\x_{t}^{(k)},\ldots, \x_{t+L-1}^{(k)}
	\end{bmatrix}.
\end{align}
Using this notion, we define the overall cost function for training the activity-detection model as
\begin{align}
	\mathcal{L}_{\mathrm{cost}}\left(\mathcal{D};\bm{\theta} \right)
	=
	\frac{1}{K}
	\sum_{k=1}^{K}
	\frac{1}{T_k-L+1}
	\sum_{t=1}^{T_k-L+1}
		\mathcal{L}\left(
			\bm{X}^{(k)}_t; \bm{\theta}
		\right).
\end{align}
In the same way, we define the anomaly score for a given sound clip $\bm{X}=\begin{bmatrix} \x_1, \ldots, \x_T \end{bmatrix}$ as
\begin{align}
	\mathcal{A}_1\left(\bm{X}; \bm{\theta}  \right)
	=
	\frac{1}{T-L+1}
	\sum_{t=1}^{T-L+1}
		\mathcal{L}\left(
			\bm{X}_t; \bm{\theta}
		\right). \label{eq:anomaly score_1}
\end{align}
%\vspace{-5pt}

\subsection{Case in which activity labels are unavailable in inference}
\label{ssec:no activity labels}
\vspace{-3pt}
When the machine activity is not automatically recorded, the anomaly score defined in \eqref{eq:anomaly score_1} cannot be computed.
%activity labels have to be annotated by hand. In this case, obtaining activity labels in the inference phase is almost impossible~(since machine condition monitoring should be performed automatically), and the anomaly score defined in \eqref{eq:anomaly score_1} cannot be computed.
To deal with this situation, we propose training an outlier detector with the embedding vectors (which are the outputs of $f$) of the training data after the activity-detection model has been trained. For example, we can use a GMM or an AE as an outlier detector. Here, the anomaly score for $\bm{X}$ is calculated as
\begin{align}
\mathcal{A}_2 \left( \bm{X}; \bm{\theta}_f, \bm{\theta}_o \right) = \frac{1}{(T-L+1)L} \sum_{t=1}^{T-L+1}\sum_{l=1}^{L} \mathcal{L}_o (\bar{\x}_l^{t}; \bm{\theta}_o), \label{eq:anomaly score_2}
\end{align}
where $\bar{\x}_l^{t}$ denotes the $l$-th embedding vector of $f(\bm{X}_t, \bm{\theta}_f)$, $\bm{\theta}_o$ denotes the parameters of the outlier detector, and $\mathcal{L}_o (\bar{\x}; \bm{\theta}_o)$ denotes the anomaly score for the outlier detector given an embedding vector $\bar{\x}$.
Here, $\bar{\x}_l^t$ is expected to be close to either $\w_1$ or $\w_2$ in \eqref{eq:classifier}. Then, if a feature vector at a time frame that includes anomalous sounds is provided to the activity-detection model, $\bar{\x}_l^t$ would be dissimilar to both vectors and the anomaly score for this embedding vector would be high. Thus, anomalous sounds will have high anomaly scores. In this way, UASD can be conducted without using activity labels in the inference phase.
We refer to this method as \textit{UASD based on outlier detection using supervised activity detection~(UASD-OD-SAD)}.

\section{EXPERIMENTS}
\label{sec:exp}
\vspace{-5pt}

\begin{table}[!t]
\caption{Requirements for activity labels.}
\vspace{-5pt}
\label{table:requirements}
	\centering
	\begin{tabular}{|l|c|c|}
		\hline
		Method & Training & Inference \\
		\hline
		(i) UASD w/ activity labels & Yes & Yes \\
		(ii) UASD-SAD & Yes & Yes \\
		(iii) UASD w/o activity labels & No & No \\
		(iv) UASD-OD-SAD & Yes & No \\
		\hline
	\end{tabular}
	\vspace{-15pt}
\end{table}

\subsection{Experimental conditions}
\label{ssec:exp conditions}
\vspace{-1pt}

To investigate the effectiveness of the proposed method under noisy conditions, we compared its performance with that of a conventional UASD method using a machine sound dataset containing environmental noise.

For evaluation, we used the slide rail dataset included in the MIMII DUE dataset~\cite{tanabe2021mimii}, as it satisfies our assumption that the input sounds contain both active and inactive sections of the target machine. Furthermore, slide rails are widely utilized in factories, and detecting their breakdown is critically important.
% Given that the proposed method is based on the assumption that the input sounds contain both active and inactive sections of the target machine, we used the slide rail dataset included in the MIMII DUE dataset~\cite{tanabe2021mimii}. 
We used the data in sections ``00'' and ``01'' in the development dataset, which contain different sounds. We annotated the active sections of the slide rail dataset for both the training and test data. To evaluate each method under low-signal-to-noise ratio~(SNR) conditions, we mixed  the original dataset with two types of environmental noise each recorded in different factories~(Factory A and B). The SNR was between $6.0$~dB and $-12.0$~dB, where the mixing procedure was the same as that previously reported~\cite{tanabe2021mimii}. Note that the original slide rail dataset already includes environmental noise, so the SNR given here is not the SNR between the clean slide rail sound and other noise. Instead, it is the SNR between the sounds of the original dataset and the additionally mixed in factory noise. The input for each method was 128-dimensional log-mel spectrograms computed with a short-time Fourier transform frame size of $64~\mathrm{ms}$ and a hop size of $50\%$.

We consider two scenarios: one in which activity labels are available in inference, and one in which they are not.
For the first scenario, we compared the proposed method, UASD-SAD, with the conventional AE-based method using the activity labels~(UASD w/ labels).
The AE for UASD w/ labels was trained and evaluated using only the data when active.
For the second scenario, we compared the proposed method, UASD-OD-SAD, with the conventional AE-based method without the activity labels~(UASD w/o labels).
The architecture of the AE was the same as reported~\cite{tanabe2021mimii}: five fully connected layers for the encoder and the decoder with batch-norm layers located between every pair of layers. 
Each input contained five consecutive frames of extracted feature vectors concatenated to a single 640-dimensional vector.
\begin{figure}[t]
	\centering
	\begin{tabular}{c}
		\begin{minipage}[t]{1.0\hsize}
			\centering
			\includegraphics[width=8cm]{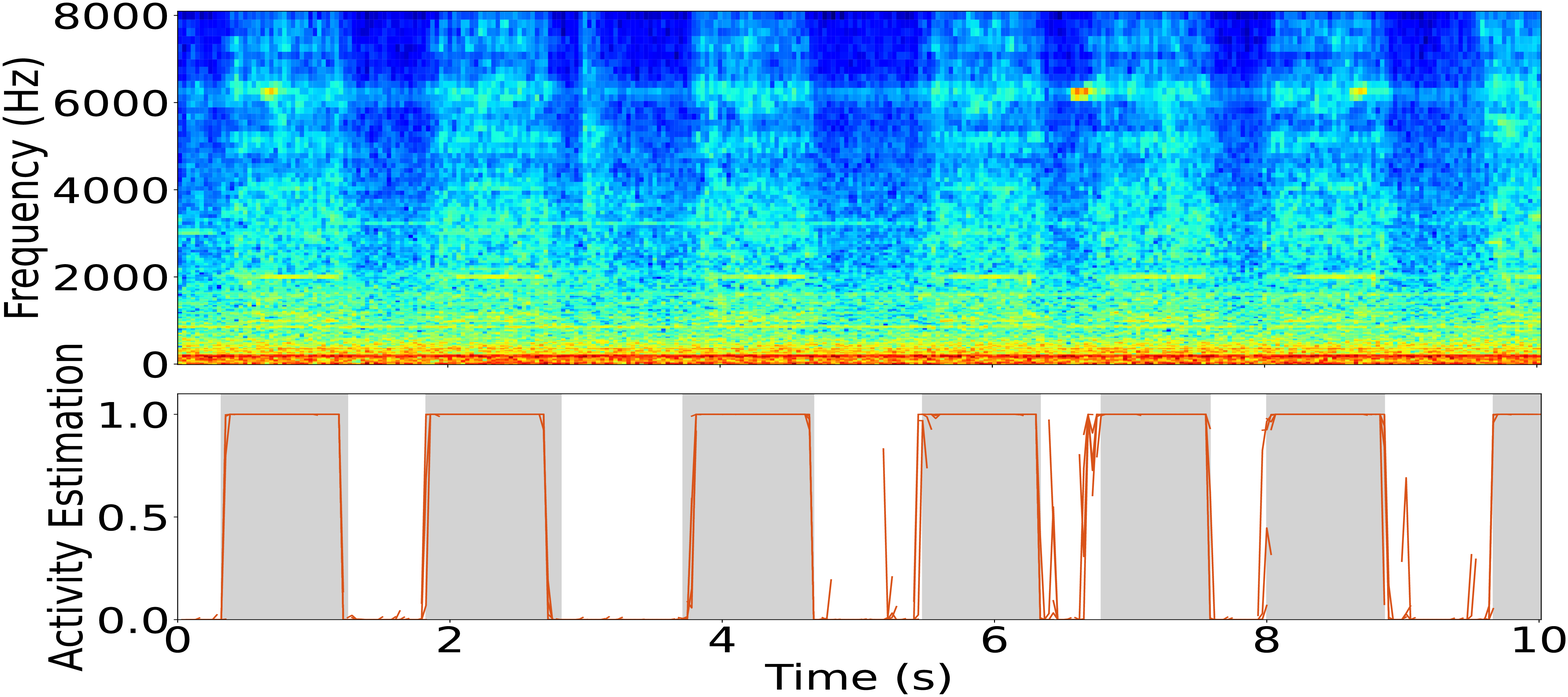}
			\subcaption{Normal sound clip}
		\end{minipage}\\
		\begin{minipage}[t]{1.0\hsize}
			\centering
			%\vspace{5pt}
			\includegraphics[width=8.0cm]{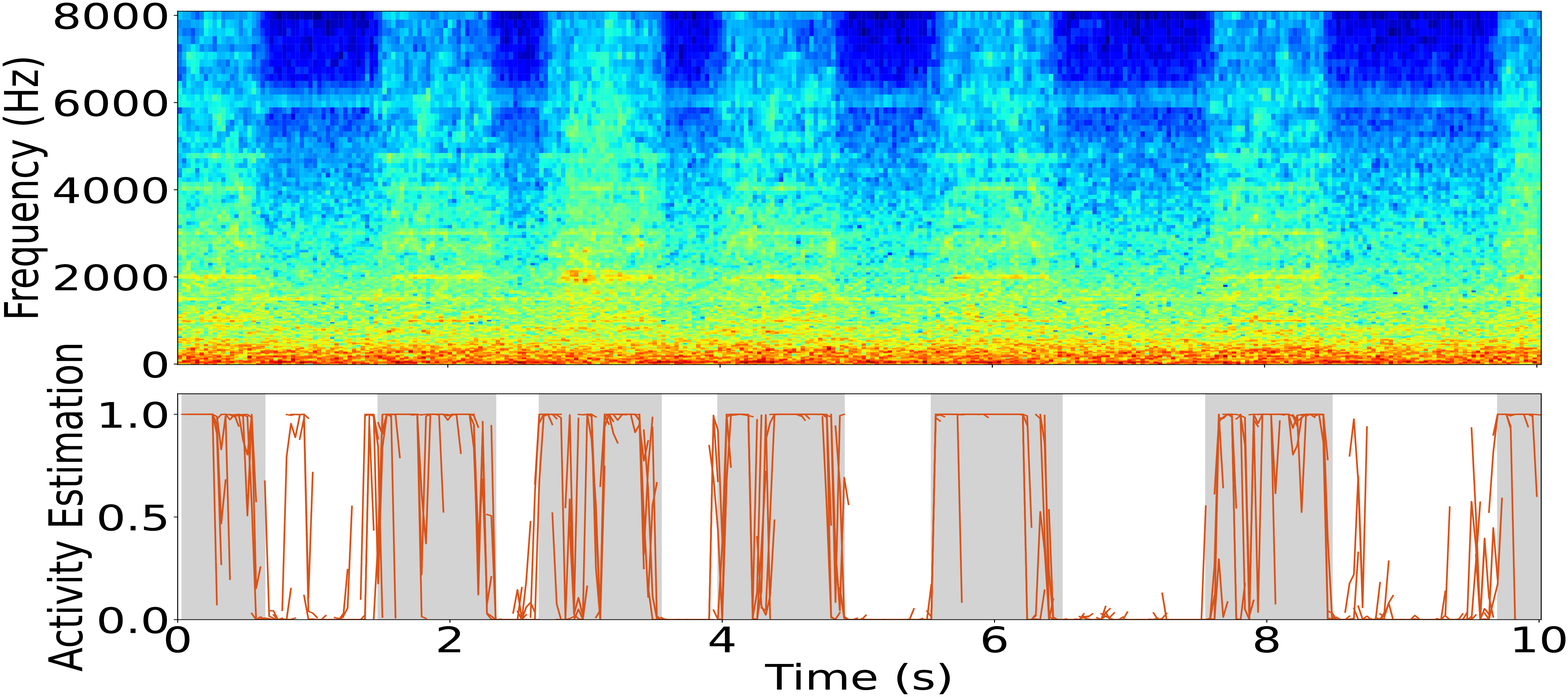}
			\subcaption{Anomalous sound clip}
		\end{minipage}
	\end{tabular}
	\caption{Example of log-mel spectrogram~(upper parts of (a),~(b)) and activity detection~(lower parts of (a),~(b)) for UASD-SAD~(section 00, Factory A noise, $\mathrm{SNR}=6.0~\mathrm{dB}$). Gray areas show ground-truth activities, and each line represents the estimated activities for $L$ consecutive frames, where $L=5$.}
	\label{fig:activity}
	\vspace{-15pt}
\end{figure}

We used a CNN-based architecture for the activity-detection model.
The feature vectors of five consecutive frames were extracted to form a two-dimensional time-frequency representation. 
The feature vectors were input to a CNN layer and then by three residual CNN blocks. 
Each residual block consisted of two CNN layers and a residual connection, where each layer had 32 channels. 
The size and stride of the convolution kernel were $3\times3$ and $1$. 
A fully connected layer then followed the CNN layers, which was applied to each feature of different time frames separately and identically. 
The architecture of this feature embedding model was designed so that the number of layers was approximately the same as the AE model, thus ensuring that the representation capacity of the two models would be close to each other. 
We used \eqref{eq:classifier} for the activity-detection model and a GMM for the outlier-detection model. 
The number of mixture components in the GMM was 5.
All neural networks were optimized by Adam~\cite{Kingma2015adam}. 
The AEs were trained for $100$ epochs, and the activity-detection models were trained for $20$ epochs. 
Note that each method has different requirements for the activity labels, as shown in Table~\ref{table:requirements}.

For the first scenario, we evaluated an ensemble of UASD w/ labels and UASD-SAD. 
For the second scenario, we performed an ensemble of UASD w/o labels and UASD-OD-SAD.
To ensemble different methods, the anomalous score of each model was first standardized and then summed up to calculate the overall anomaly score.
The standardization of the anomaly score $\mathcal{A}(\bm{X})$ of a test data $\bm{X}$ was conducted by
\begin{align}
\tilde{\mathcal{A}}(\bm{X}) = \left(\mathcal{A}(\bm{X}) - \mu \right)/\sqrt{\sigma^2 + \epsilon},
\end{align}
where $\mu$ and $\sigma^2$ are the mean and the variance of $\mathcal{A}(\bm{X})$ of the training data, respectively, and $\epsilon$ is a constant positive value. We used \mbox{$\epsilon=1000$} for UASD-SAD since the variance of the anomaly score for the evaluation data tended to be much larger than $\sigma^2$. We used \mbox{$\epsilon=0$} for all the other methods.

\subsection{Results}
\label{ssec: results}
\vspace{-5pt}
%\vspace{-5pt}
\begin{table}
\caption{AUC values for anomaly detection (\%).}
\label{table:AUC}
\vspace{-8pt}
	\begin{subtable}{0.48\textwidth}
	\centering
		%\vspace{7pt}
		\caption{Methods that require activity labels in inference.}
		\vspace{-3pt}
		\resizebox{1\textwidth}{!}{
		\begin{tabular}{@{}lc|cccccccc@{}}
			\hline
			 \multicolumn{2}{r|}{Noise type}  & \multicolumn{4}{c|}{Factory A}   & \multicolumn{4}{c}{Factory B}\\
			Method & \multicolumn{1}{r|}{SNR~[dB]}   & 6.0 & 0 & -6.0 & -12.0 & 6.0 & 0 & -6.0 & -12.0\\
			\hline \hline
			\multicolumn{10}{c}{\textbf{Section 00}}\\
			\hline
			\multicolumn{2}{@{}l|}{(i) UASD w/ labels}     & 76.6& 70.8 & 59.5 & 50.7 & 73.4 & 66.3 & 58.6 & 51.0\\
			\multicolumn{2}{@{}l|}{(ii) UASD-SAD}            & 73.0 & 70.7 & \textbf{62.8} & \textbf{58.7} & 68.9 & 68.9 & \textbf{68.8} & \textbf{55.1}\\
			\multicolumn{2}{@{}l|}{Ensemble of (i) and (ii)}  & \textbf{81.5} & \textbf{73.6} & 58.3 & 51.5 & \textbf{77.7} & \textbf{72.1} & 62.7 & 52.1\\
			\hline
			\multicolumn{10}{c}{\textbf{Section 01}}\\
			\hline
			\multicolumn{2}{@{}l|}{(i) UASD w/ labels}     & \textbf{82.2} & 77.7 & 72.4 & \textbf{63.3} & \textbf{82.3} & 76.7 & 67.8 & \textbf{60.5} \\
			\multicolumn{2}{@{}l|}{(ii) UASD-SAD}            & 81.6 & 73.9 & 63.5 & 53.7 & 80.0 & 74.3 & 64.2 & 53.2\\
			\multicolumn{2}{@{}l|}{Ensemble of (i) and (ii)}            & 81.0 & \textbf{81.5} & \textbf{74.9} & 58.5 & 82.1 & \textbf{82.6} & \textbf{72.3} & 57.5 \\
			\hline
			\multicolumn{10}{c}{\textbf{Average}}\\
			\hline
			\multicolumn{2}{@{}l|}{(i) UASD w/ labels}     & 79.4 & 74.3 & 66.0 & \textbf{57.0} & 77.9 & 71.5 & 63.2 & \textbf{55.8} \\
			\multicolumn{2}{@{}l|}{(ii) UASD-SAD}            & 77.3 & 72.3 & 63.2 & 56.2 & 74.5 & 71.6 & 66.5 & 54.2\\
			\multicolumn{2}{@{}l|}{Ensemble of (i) and (ii)}            & \textbf{81.3} & \textbf{77.6} & \textbf{66.6} & 55.0 & \textbf{79.9} & \textbf{77.4} & \textbf{67.5} & 54.8 \\
			\hline
		\end{tabular}}
	\end{subtable}
	\begin{subtable}{0.48\textwidth}
	\centering
        \vspace{7pt}
		\caption{Methods that do not require activity labels in inference.}
		\vspace{-3pt}
		\resizebox{1\textwidth}{!}{
		\begin{tabular}{@{}lc|cccccccc@{}}
			\hline
			\multicolumn{2}{r|}{Noise type}  & \multicolumn{4}{c|}{Factory A}   & \multicolumn{4}{c}{Factory B}\\
			Method & \multicolumn{1}{r|}{SNR~[dB]}   & 6.0 & 0 & -6.0 & -12.0 & 6.0 & 0 & -6.0 & -12.0 \\
			\hline \hline
			\multicolumn{10}{c}{\textbf{Section 00}}\\
			\hline
			\multicolumn{2}{@{}l|}{(iii) UASD w/o labels}     & 70.3 & 64.9 & 55.8 & 51.7 & 70.2 & 61.5 & 55.4 & 51.2\\
			\multicolumn{2}{@{}l|}{(iv) UASD-OD-SAD}    & 76.2 & 66.5 & 56.3 & 52.2 & 70.1 & \textbf{68.7}& \textbf{64.7} & \textbf{51.7}\\
			\multicolumn{2}{@{}l|}{Ensemble of (iii) and (iv)}              & \textbf{77.5} & \textbf{68.3} & \textbf{57.0} & \textbf{52.3} & \textbf{74.6} & 67.7 & 60.7 & \textbf{51.7}\\
			\hline
			\multicolumn{10}{c}{\textbf{Section 01}}\\
			\hline
			\multicolumn{2}{ @{}l |}{(iii) UASD w/o labels}     & 80.2 & 74.0 & 65.9 & \textbf{57.4} & 79.5 & 72.0 & 63.8 & \textbf{57.0}\\
			\multicolumn{2}{ @{}l |}{(iv) UASD-OD-SAD}  &  71.4 & 72.6 & 57.4 & 47.1 & 72.1 &69.1 & 62.1 & 47.5\\
			\multicolumn{2}{ @{}l |}{Ensemble of (iii) and (iv)}             & \textbf{80.5} & \textbf{78.3} & \textbf{68.2} & 53.9 & \textbf{81.0} & \textbf{78.1} &\textbf{68.2} & 53.7\\
			\hline
			\multicolumn{10}{c}{\textbf{Average}}\\
			\hline
			\multicolumn{2}{ @{}l |}{(iii) UASD w/o labels}     & 75.3 & 69.5 & 60.9 & \textbf{54.6} & 74.9 & 66.8 & 59.6 & \textbf{53.8} \\
			\multicolumn{2}{ @{}l |}{(iv) UASD-OD-SAD}  &  73.8 & 69.6 & 56.9 & 49.7 & 71.1 & 68.9 & 63.4 & 49.6\\
			\multicolumn{2}{ @{}l |}{Ensemble of (iii) and (iv)}             & \textbf{79.0} & \textbf{73.3} & \textbf{62.6} & 53.1 & \textbf{77.8} & \textbf{72.9} &\textbf{64.5} & 52.7\\
			\hline
		\end{tabular}}
	\end{subtable}
	%\vspace{10pt}
	\vspace{-5pt}
\end{table}

Examples of the activity detection of UASD-SAD are provided in Fig.~\ref{fig:activity}. 
As we can see, activity was detected correctly for normal sounds, while there were many errors in activity detection for anomalous sounds.
Therefore, the proposed method is expected to detect anomalies based on the error of activity detection.

Table~\ref{table:AUC} shows the area under the receiver operating characteristic curve (AUC) for anomaly-detection performance.
First, in most conditions, the methods that require the activity labels in the inference phase had higher AUC values than those that do not require the activity labels. 
Next, in many of the noise conditions in Section 00, UASD-SAD showed higher AUC values than UASD w/ label, but in Section 01, UASD w/ label showed higher AUC values.
Also, in most of the noise conditions in Section 00, UASD-OD-SAD showed higher AUC values than UASD w/o label, but in Section 01, UASD w/o label showed higher AUC values.
One of the reasons for the conflicting results obtained in Section 00 and Section 01 is most likely the difference of the similarity between the target machine sound and the environmental noise. 
The proposed method is expected to show high performance when the target machine sound is somewhat similar to the noise. 
On the other hand, when the target machine sound is not similar to the noise at all, anomaly detection cannot be performed well because the auxiliary task, activity detection, is too easy. 
In fact, the noise of factories A and B contained sounds similar to the slide rail in Section 00 but not to the slide rail in Section 01. Overall, the proposed methods achieved a better performance than the conventional methods when the noise was similar to the target machine sound, which is a condition that degraded the performance of the conventional methods substantially. This advantage is crucial in practical situations, since factories often run several similar machines in the same area. In this case, environmental noise tends to be similar to the target machine sound.

Also, the results showed that the ensembles achieved higher AUC values than the conventional methods for both sections 00 and 01, except for the SNR of $-12.0$~dB. These results indicate that the proposed method is also useful for improving the anomaly-detection performance of the conventional methods complementarily by means of an ensemble. It is also suggested that the proposed method does not contribute to the performance improvement at all when the SNR is extremely low.

\section{CONCLUSION}
\label{sec:conclusion}
\vspace{-5pt}

We proposed a method for anomalous sound detection based on machine activity detection. 
The proposed method calculates the anomaly score based on the error of activity detection if the ground-truth activity labels are available in the inference phase. 
If these labels are not available, it performs outlier detection for the embeddings obtained in the activity-detection model. The experimental results indicate that the proposed method achieves a better performance than the conventional method particularly when the environmental noise contains sounds similar to the target machine sound, which is a crucial advantage in practical applications.
In addition, the proposed method improved the anomaly-detection performance of the conventional method complementarily by means of an ensemble.
%The experimental results also indicate that if the SNR is too low, the effect of the proposed method is small or the proposed method conversely reduces the anomaly-detection performance. 
% We will study how to automatically control the balance between the proposed method and the conventional method by SNR in future work.

%\vfill\pagebreak
\bibliographystyle{IEEEbib}
\bibliography{str_def_abrv, refs}

\end{document}